\documentclass[11pt]{article}
\usepackage{amssymb,setspace}

\begin{document}

\bibliographystyle{unsrt}

\title{A modified-Lorentz-Transformation based \\ gravity model confirming basic GRT experiments.}

\date{}
\author{ Jan Broekaert\footnote{CLEA-FUND, Vrije Universiteit Brussel,  Belgium}} 
\maketitle

\begin{abstract}
\noindent
Implementing Poincar\'e's \emph{geometric conventionalism} a scalar Lorentz-covariant gravity model is obtained based on gravitationally modified Lorentz transformations (or GMLT).  The modification essentially  consists of an appropriate space-time and momentum-energy scaling (``normalization") relative to a nondynamical flat background geometry according to an isotropic, nonsingular gravitational \emph{affecting} function $\Phi(\mathbf{r})$.   Elimination of the gravitationally \emph{unaffected} $S_0$ perspective by local composition of  space-time GMLT recovers the local  Minkowskian  metric and thus preserves  the invariance of the locally observed velocity of light. The associated energy-momentum GMLT   provides  a covariant Hamiltonian description for test particles and  photons which, in a static gravitational field configuration,  endorses  the four  `basic' experiments for testing  General Relativity Theory: gravitational {\it i)} deflection of light, {\it ii)} precession of perihelia, {\it iii)} delay of radar echo, {\it iv)} shift of spectral  lines. The model recovers the  Lagrangian of the  Lorentz-Poincar\'e gravity model  by Torgny Sj\"odin and integrates elements of  the precursor gravitational theories, with spatially Variable Speed of Light (VSL) by Einstein  and Abraham, and gravitationally variable mass by  Nordstr\"om.   \end{abstract}

\vspace{22 pt}


\section{Introduction}


It is well known that it is possible to treat gravitation in a flat, ``unrenormalized" pseudo-Euclidean
space-time,  \emph{i.e} the one used in special relativity. Then, because of the universal action of gravity, even the meter sticks and and the clocks are affected by gravity. The ``renormalized" space-time observed by those modified rods and clocks turns out to be Riemannian. By the standard treatment of the field theory, \emph{i.e.},  assumption of the existence of  a potential, Lagrangian with minimal coupling, and variational principle, both the field equations and the equations of motion are derived.  The latter ones are in general not the Euler-Lagrange equations but the ones of Infeld-Khalatnikov-Kalman \cite{Kalman1961FoPh} generalized to cope with Finsler spaces by Cavalleri and Spinelli \cite{CavalleriSpinelli1974FoPh}. The agreement with observations depends on the tensor rank of the potential. Actually, a \emph{scalar} theory (\emph{i.e.}, a theory whose potential is a zero-rank tensor) gives no deviation of the light ray, and a perihelion precession $-1/6$ the correct value \cite{Laue1917FoPh}. A \emph{vector} theory  (\emph{i.e.}, a theory whose potential is a first-rank tensor) has to be immediately discarded since in an attractive theory ---as gravitation--- it leads to self--acceleration, moreover  it violates the Weak Equivalence Principle  \cite{WhitrowMorduch1965}. A \emph{second rank} gravitational theory has  been  initiated by Fierz \cite{Fierz1939FoPh}, was subsequently developed by many authors \cite{fieldtheorygravDeser, Deser1970FoPh} and shown to be valid even with the most general gauge by  Cavalleri and Spinelli  \cite{CavalleriSpinelliCombi1975FoPh}. The corresponding ``renormalized" space-time is Riemannian and the theory, obtained by an iterative procedure, converges to Einstein's theory. From the above point of view, any new scalar gravitational theory seems to be discarded. However, what is proposed here, is a modified scalar theory leading to the same predictions derivable from the PPN approximation \cite{Broekaert2004a, Broekaert2004b}, i.e. to the expansion of Einstein's theory into flat space but truncated after the second order in $\kappa/r$.\\
However this model differs from Einstein's theory; e.g. it does not lead to ``black holes", in the sense that the speed of light decreases with $r$ but vanishes only for $r \to 0$. The divergence for finite $r$ ---considered as a failure by Einstein himself--- is not  present in it.  Moreover there is some hint from astronomy, that the $\gamma$-ray bursts could be explained by the collision of a binary neutron star system, but each with a mass $>5 M_{Sol}$ as required for the observed energy emission.  \emph{I.e.} well beyond the limit of collapse to a black hole according to GRT (\footnote{Private communication by G. Cavalleri.}). It is therefore  worthwhile to develop in particular, a nonsingular,  modified scalar theory whose predictions are in agreement with present gravitational experiments.\\
The `modified' scalar gravity model is inspired by the principle of gravitation as a  physical affecting of space and time measurement rather than intrinsic geometric modification \cite{Broekaert2002FoPh}.   The effectiveness of the model relies on Poincar\'e's conventionalism of geometry. Poincar\'e's concern was the complementing nature of geometry and physics (optics), leading to a theoretical  indetermination which precludes the extraction of the  unique fundamental geometry of nature. Poincar\'e's well known argument states that a physical theory is a set of laws which are a combined result of  geometry (`G') and physical rules (`P'), and, this inextricable combination (`G+P')  leaves open  to convention the nature of the intrinsic basic geometry \cite{conventionalismgrouped}. \\
Various attempts to model gravitation in the Lorentz-Poincar\'e sense ---i.e. with Euclidean space and time and physical effects  like `rod contraction and clock dilation'--- have been made (cf. references in \cite{PocciSjodin1980FoPh,Arminjon2004aFoPh}, and Thirring's ``normalization" \cite{ThirringW1961FoPh}), most often invoking spatially variable speed of light (compare to recent  VSL-theories for epochal variation of universal constants in cosmology, e.g. in the work of J.W. Moffat \cite{Moffat1993FoPh},  and A. Albrecht and J. Magueijo  \cite{AlbrechtMagueijo1999FoPh}).
Historically the main precursor  theories  of gravitation  were conceived using a gravitationally dependent velocity of light (e.g.,  Einstein \cite{Einstein19071912grouped} and Abraham \cite{Abraham1912grouped}) or mass (Nordstr\"om \cite{Nordstrom1913grouped}).  The ineffectiveness of the approach ---the manifest absence of  Lorentz symmetry--- was subsequently recognized (see e.g. Norton's overview \cite{Norton1992FoPh}).\\
The present model uses  a spatially variable speed of light but establishes the local Lorentz symmetry due the introduction of scaling functions on measured space and time intervals.   The model recovers the particle Lagrangian  of Torgny Sj\"odin's  Lorentz-Poincar\'e type gravity model [22-24]\nocite{Sjodin1982FoPh, Sjodin1990FoPh, PodlahaSjodin1984FoPh}, includes the VSL-property of the  modified Lorentz transformations written by Abraham \cite{Abraham1912grouped} in continuation  of related developments by Einstein \cite{Einstein19071912grouped},  and aspects of  mass variability according Nordstr\"om's scalar gravitation theory \cite{Nordstrom1913grouped}.\\
The gravity model ---in the present paper only developed for test particles in a static spherically symmetric configuration--- does endorse  the four `basic' GRT experiments to required order in the Schwarzschild radius $\kappa$ while retaining a Lorentz-Poincar\'e interpretation. \\
In the following Section (2) we develop the concept of gravitationally modified Lorentz  transformation (GMLT). Modifications, or generalizations, of the Lorentz transformations ---introduced previously for different purposes in the literature, e.g. \cite{OtherMLTgrouped}--- break  Lorentz covariance in some limit, e.g. at small energy scale.  In these limits the modified transformations  lack the Lorentz symmetry but still relate quantities as observed by different observers. In the present model the proposed modifications are due to gravitation and aim at an adequate description of relativistic gravitational effects.  The particle and photon  Hamiltonian, derived by GMLT, are rendered by applying the Newtonian fit. These expressions allow one to compare the present scalar  gravitation model with the \emph{deficient} classical  scalar and vectorial models, as outlined by Cavalleri and Spinelli \cite{CavalleriSpinelli1983aFoPh}.\\
Section (3) presents some   applications and consequences of these  transformations, in particular  calculations of the `basic' GRT experiments and kinematic modification of gravitational acceleration.  The  succesfull description of  the `basic' experiments are  evidently  not sufficient for considering the present model as yet a full alternative metrization of GRT. We refer to recent work for some further development of the model concerning dynamic sources \cite{Broekaert2004a} and extended test masses \cite{Broekaert2004b} in line with GRT.  \\
We remark the recent scientific literature accounts of numerous ---in aspects--- related approaches to our model, e.g. in the sense of either: VSL-models \cite{Magueijo2003FoPh, AlbrechtMagueijo1999FoPh, Moffat1993FoPh}, polarizable vacuum models \cite{Puthoff2002FoPh}, scalar field gravitation models \cite{Arminjon2004aFoPh, scalarfieldmodelsgrouped}, nonsingular ether gravitation models \cite{CavalleriTonni1997FoPh} and, field conceptions of gravitation [31-33]\nocite{Weinberg1965FoPh, CavalleriSpinelli1980FoPh, PittsSchieve2004FoPh}.

\section{Gravitationally modified Lorentz transformations}
Following the geometric conventionalist approach, we  introduce a gravitational affecting of observed space and time intervals (`rod intervals' and `clock intervals') in a gravitational field, \emph{i.e.} the  physical \emph{congruence standard}. We let observed space and time  intervals, and the speed of light be heterogeneously and isotropically affected.   The gravitational
affecting at position ${\bf r}$ is given by the function ${\Phi \left( {\bf r} \right)}$,
$0\leq{\Phi   \left( {\bf r} \right)}\leq 1$, as the ratio of  a space interval observed by the gravitationally unaffected
{$S_0$} with {\em coordinate} geometry, and the same infinitesimal interval as observed by gravitationally
affected observer {$S'$} with {\em natural} geometry, both at ${\bf r}$.
\begin{eqnarray} 
 d {\bf x} = d {\bf x}' {\Phi   \left( {\bf r} \right)}   &,&  d  t  = \frac{d t'}{ {\Phi   \left( {\bf r} \right)}}, \label{staticrelation}
\end{eqnarray}
where the inverse scaling effect is set on the relation of affected and unaffected  infinitesimal time intervals.  Within the scope of the present model there is no need to attribute to this time ratio a separate affecting function (compare \cite{Sjodin1990FoPh}). The invariance of the \emph{locally observed}  velocity of light is secured by adequately balancing the variability of speed of light and gravitational affecting of space and time measurement:
\begin{eqnarray} 
 {c ({\bf r})} &=& {c '} {\Phi   \left( {\bf r} \right)}^2. \label{VSL}
\end{eqnarray}   
The imposed ratio of velocities of light  $c({\bf r})$ relative to  gravitationally unaffected {$S_0$},  and $c'$ as observed by  affected observer  {$S'$}, is consistent with affected space and time ratios. 
The localy observed velocity of light `in vacuum'  remains the usual universal constant ${c'}$.  We emphasize that relative to the  gravitationally unaffected observer {$S_0$} we have posited an explicitly  variable speed of light,  and explicit  `observed' time dilation and observed space interval contraction. All observed variables, $\{d {\bf x}', d {\bf t}', c' \}$, are primed and refer to the affected observer $S'$, while all variables $\{d {\bf x}, d {\bf t}, c  \}$ without prime refer to the `unphysical' perspective of the unaffected $S_0$. If the affecting function $\Phi({\bf r})$ is known, the latter  variables  can  immediately be related to those of $S'$ by calculation. Actually expressions  should refer to the physical $S'$ perspective. Still, comparison with GRT's results is done most often in the coordinate perspective of $S_0$, which does not necessitate a
protocol for  measurement of extensive length.  Finally we note that, the general case of relations Eq. (\ref{staticrelation})
with $\Phi = \Phi ({\bf r},t)$ are equally valid with respect to the gravitational affecting principle. However, the static field configuration is sufficient for the scope of the present paper. 

Subsequently we introduce the effect of kinematics on observed space and time intervals in a given static field ${\Phi   \left( {\bf r} \right)}$.

\subsection{Space and Time GMLT}
In general a purely gravitational affecting Eq. (\ref{staticrelation}) and {\em kinematic} affecting are required in the observer transformations. In the Lorentz-Poincar\'e interpretation the latter consists of, \emph{
i)} physical longitudinal Lorentz contraction of a `material rod' by a factor $\gamma ({v})^{-1}$,  \emph{ii)}  physical
Lorentz  time dilation of a moving `clock' by a factor $\gamma ({v})$, both with  velocity
${\bf v}$ relative to the field $\Phi$. The composition of both effects ---gravitational and kinematic--- and {\it standard synchronization} or {\it Einstein synchronization} [34-37] \nocite{MansouriSexl1977FoPh, Sjodin1979FoPh, CavalleriSpinelli1983bFoPh, CavalleriBernasconi1989FoPh} into an observer transformation between momentary locally coincident observers {$S'$} and {$S_0$}, leads directly to the gravitationally modified Lorentz transformation, the GMLT  of 
{$S_0$} to  {$S'$}:
\begin{eqnarray}
d{\bf x}' &=& \left(( {d{\bf x}_\parallel} -{\bf u}  d t)  {\gamma  \left( u \right)}     +
  {d{\bf x}_\perp} \right) \frac{1}{{\Phi   \left( {\bf r} \right)}},  \label{spacesotosac} \\ 
d t' &=& \left( d t - \frac{ {\bf u}. d{\bf x} }{{c ({\bf r})}^2}  \right)  {\gamma  \left( u \right)}  {\Phi   \left( {\bf r} \right)},
\label{timesotosac}
\end{eqnarray} 
and the inverse  GMLT {$S'$} to {$S_0$} :
 \begin{eqnarray}
d{\bf x} &=& \left(( {d{\bf x}_\parallel}' -{{\bf u}'}\ d t')  {\gamma  \left( u' \right)}     +
  {d{\bf x}_\perp}' \right) {\Phi   \left( {\bf r} \right)} , \label{spacesactoso}   \\ 
d t &=& \left( d t' - \frac{ {{\bf u}'}. d{\bf x}' }{c'^2}  \right) \frac{ {\gamma  \left( u' \right)}}{{\Phi   \left( {\bf r} \right)}}, 
\label{timesactoso}
\end{eqnarray}  
with ${c ({\bf r})}$ as in Eq. (\ref{VSL}), and ${\gamma  \left( u \right)}$ and ${\bf u}$ satisfying:
\begin{eqnarray}
{\bf u} &=& - {{\bf u}'} {\Phi \left( r \right)}^2,  \\
{\gamma  \left( u' \right)} &=& \left(1- \left(\frac{{u'}}{{c '}}\right)^2\right)^{-1/2} =\left(1- \left(\frac{u}{c ({\bf r})}\right)^2\right)^{-1/2}.
\end{eqnarray}
From which the usual velocity relation, angle conversion, and $\gamma$ relations are obtained, {$S_0$} to {$S'$}:
\begin{eqnarray}
{{\bf v}'} &=& \frac{{{\bf v}_\parallel} - {\bf u} + {\gamma  \left( u \right)}^{-1} {\bf v}_\perp  }{\left( 1 -
 {\bf u}.{\bf v} {c ({\bf r})}^{-2}\right) {\Phi   \left( {\bf r} \right)}^2} , \ \ \  {\bf v} =
\frac{d {\bf x}}{dt} \ , \ {{\bf v}'} =
\frac{d {\bf x}'}{dt'},  \label{velocityrelation} \\ 
\tan \theta' &=& - \frac{\tan \theta}{{\gamma  \left( u \right)}  \left( 1 -  u {v_\parallel}^{-1} \right)}, \ \ \  \theta =
\angle({\bf u}, d {\bf x}) ,\ \ \ \theta' = \angle({{\bf u}'},d {\bf x}'), \\
{\gamma  \left( v' \right)}  &=&    {\gamma  \left( v\right)}  {\gamma  \left( u \right)}     \left(1 -  {\bf u}  .{\bf v}    {c (r)}^{-2}  \right)  . \label{gammarelation}
\end{eqnarray}
This  first type of  GMLT Eqs. (\ref{spacesactoso},  \ref{timesactoso}) or eqs. (\ref{spacesotosac},  \ref{timesotosac}) relate  affected $S'$ and  unaffected $S_0$ observers, and therefore need not  constitute a Lorentz  group. These transformation are in fact related to the transformations used in early gravitation models by Abraham   \cite{Abraham1912grouped} which elaborated  the concept of `light velocity as gravitational potential' introduced, and subsequently rejected on grounds of relativistic incompatibility, by Einstein   \cite{Einstein19071912grouped}. In their approach the modified transformations related physical, thus  affected, observers.   In the present model  Lorentz symmetry is  maintained between \emph{affected} locally coincident observers, {\em i.e.} $\Phi_1 = \Phi_2$, for then  the  composition of two GMLT's  $S'_1$ to $S_0$   eqs. (\ref{spacesactoso}, \ref{timesactoso}) and  $S_0$ to $S'_2$ eqs. (\ref{spacesotosac}, \ref{timesotosac}),  by elimination of their mutual  reference to {$S_0$}, leads to a  regular Lorentz transformation.  \\
In the general case the space time relations between affected observers lead to the introduction of  the second type of space-time modified Lorentz transformations. These GMLT relate two distinctly, gravitationally and kinematically, affected observers $S'_1$ and $S'_2$, $\Phi_1 \neq \Phi_2$,  ${\mathbf u_1} \not  \parallel {\mathbf u_2}$.  These are again obtained by elimination of the {$S_0$} perspective from a composition of two GMLT's  eqs. (\ref{spacesactoso},  \ref{timesactoso})  and   eqs. (\ref{spacesotosac}, \ref{timesotosac}). In the Appendix these GMLT expressions are given taking into account that  no kinematical reference should be made relative to the ``preferred" frame of $S_0$ .  
The presence of scaling factors ${\Phi_1}/{\Phi_2}$ and distinct speeds  of light $c_1=c (\mathbf{r}_1)$ and $c_2=c (\mathbf{r}_2)$ in the resulting transformation $S'_1$ to $S'_2$  prevent one to obtain the Lorentz symmetry (e.g. \cite{Rindler1979}, sec. 2.17). However, in a gravitational field the Lorentz symmetry needs only  be  regained in the local limit  $\Phi_1 \to \Phi_2$, which requirement is satisfied (Appendix eqs. (\ref{spacesactosac},  \ref{timesactosac}). In fact  all expressions ---relative frame velocities, kinematical contraction factors, Thomas angle--- are verified to smoothly recover the Special Relativistic expressions in the local limit.

Standard literature on GRT describes  celestial mechanical problems in the coordinate space perspective $S_0$. Comparison of present results with GRT will therefore not require  a GMLT between affected observers. Even, most problems are solved more conveniently relative to {$S_0$}, and if necessary results can be transformed into any affected perspective {$S'$} afterwards.

\subsection{Energy and Momentum GMLT}
The presence of the scaling function and spatial variability of the speed of light in the space-time modified Lorentz transformations $S'_0$ to $S'$ eqs. (\ref{spacesotosac},  \ref{timesotosac}) impedes the immediate transcription into  momentum-energy modified Lorentz transformations.
In order to derive the momentum-energy GMLT, we multiply  the sides of the space-time GMLT  Eq. (\ref{spacesotosac}) and  Eq. (\ref{timesotosac}) by
\begin{eqnarray}
{m_0'} {\gamma  \left( v' \right)}\frac{1}{d t'} &=& {m_0'} \frac{{\gamma  \left( v\right)}}{{\Phi   \left( {\bf r} \right)}}\frac{1}{d t},   \label{timedivider}
\end{eqnarray}
where ${m_0'} $  is the rest mass of a test body as attributed by an affected locally coincident observer $S'$. With an additional factor  ${c '}^2$  Eq. (\ref{timesotosac}) and consistent  reordering into the Lorentz transformation \emph{form}, we obtain:
\begin{eqnarray}
 {m_0'} {\gamma  \left( v' \right)} {{\bf v}'}  & =&      \left(\left(m_0' \frac{{\gamma  \left( v\right)}}{{\Phi   \left( {\bf r} \right)}^{3+\delta}} {{\bf v}_\parallel} - m_0' \frac{{\gamma  \left( v\right)} }{{\Phi   \left( {\bf r} \right)}^{3+\delta}}c ({\bf r})^2 \frac{{\bf u} }{{c ({\bf r})}^2}  \right)  {\gamma_u}     + {m_0'} \frac{ {\gamma  \left( v\right)}}{{\Phi   \left( {\bf r} \right)}^{3+\delta}}{{\bf v}_\perp} \right)  {\Phi   \left( {\bf r} \right)^{1+\delta}},     \label{proto1} \\ 
{m_0'} {\gamma  \left( v' \right)} {c '}^2    & =  &    \left( {m_0'}\frac{{\gamma  \left( v\right)} }{{\Phi  
\left( {\bf r} \right)}^{3+\delta}} {c ({\bf r})}^2  - {\bf u}. {m_0'} \frac{{\gamma  \left( v\right)}}{{\Phi   \left( {\bf r} \right)}^{3+\delta}} {\bf
v}\right)   {\gamma_u} \frac{1}{\Phi   \left( {\bf r} \right)^{1-\delta}},   \label{proto2}  
\end{eqnarray} 
where $\delta$ is an as yet to be fixed  numerical parameter. Equations  (\ref{proto1}) and  (\ref{proto2}) lead to a  consistent interpretation of  the test body's relative energy and momentum by an affected observer {$S'$}, locally coincident with the test body and attributing to it a relative velocity ${\bf v}'$,  according: 
\begin{eqnarray} 
{{\bf p}'} \equiv {m_0'} {\gamma \left( v' \right)} {{\bf v}'} , & & \ \ \ {E'} \equiv {m_0'}  {\gamma  \left( v'
\right)} {c '}^2,
\end{eqnarray}  
and the  corresponding  momentum and energy as attributed by $S_0$: 
\begin{eqnarray}
{{\bf p}} \ \equiv \ {m ({\bf r}, v)} {\bf v}, & & \ \ \ E  \ \equiv \  {m ({\bf r}, v)} {c ({\bf r})}^2, 
\label{momentumandenergy}
\end{eqnarray}
with location and velocity dependent mass ${m ({\bf r}, v)}$ in the $S_0$ perspective:
\begin{eqnarray}
 {m ({\bf r}, v)} &\equiv&  \  {m_0'} \frac{{\gamma  \left( v \right)}}{{\Phi   \left( {\bf r} \right)}^{3+\delta}}.  \label{mass}
\end{eqnarray}
The momentum-energy GMLT eqs. (\ref{proto1}, \ref{proto2}) are  \emph{a priori} not unique as the powers of the factorized affecting functions depend on the free numerical parameter $\delta$. However, by fitting the model to the Newtonian limit and experiments, this adaptable parameter is fixed at
\begin{eqnarray} 
\delta &=& 0.   \label{delta=0} 
\end{eqnarray}
We will therefore  adopt GMLT eqs. (\ref{proto1}, \ref{proto2})   immediately with  Eq. (\ref{delta=0})  and only mention the effect of the adaptable parameter $\delta$ again when appropriate.\\
The location dependence of mass Eq. (\ref{mass}) relative to $S_0$, considering Eq. (\ref{delta=0}), corresponds  qualitatively  to Mach's principle of mass induction: mass increases when approaching the gravitational source ($\Phi  \to \Phi_{min.} $), but remains finite on asymptotic separation $m_\infty= {m_0'}$ ($\Phi_\infty =1$).\\ 
The fundamental expression, relative to $S_0$, for the energy  $E = E({\bf r}, {\bf p})$ of a test body with rest mass $m_0$  in a gravitational field $\Phi$  is obtained by elimination of the velocity ${\bf v}$ in Eq. (\ref{momentumandenergy}):
\begin{eqnarray}
 E^2 - {c ({\bf r})}^2 p^2   &=&  {m_0 ({\bf r})}^2  c^4,   \label{massHamiltonian}
\end{eqnarray}
which is  the formal equivalent of the energy relation for a free mass in SRT.  By extension, in classical light approximation $ m_0 \equiv 0$, the energy of a photon in a gravitational field is given by: 
\begin{eqnarray}
   E  & = & p {c ({\bf r})}.   \label{lightHamiltonian}
\end{eqnarray}

With the definitions of momentum Eq. (\ref{momentumandenergy} a),  energy Eq. (\ref{momentumandenergy} b), and mass Eq. (\ref{mass}), in
the unaffected perspective of $S_0$ the momentum-energy GMLT {$S_0$} to {$S'$} becomes:
\begin{eqnarray}
{{\bf p}'} &=&  \left(\left( {{\bf p}}_\parallel  -   \frac{E}{ {c ({\bf r})}^2} {\bf u} \right)   {\gamma  \left( u
\right)}   + {{\bf p}}_\perp \right) {\Phi   \left( {\bf r} \right)},   \label{momentumsotosac}\\
{E'} &=& \left( E - {{\bf p}} . {\bf u}    \right)  \frac{{\gamma  \left( u \right)}}{{\Phi   \left( {\bf r} \right)}},
 \label{energysotosac}
\end{eqnarray}
while the inverse transformation  {$S'$} to {$S_0$} is given by:
\begin{eqnarray}
{{\bf p}} &=&  \left(\left( {{\bf p}'_\parallel} - \frac{{E'}}{ {c '}^2}  {{\bf u}'} \right) {\gamma  \left( u' \right)}  
+ {{\bf p}'_\perp}  \right) \frac{1}{{\Phi \left( {\bf r} \right)}},  \label{momentumsactoso}\\
E &=& \left( {E'} - {{\bf p}'} . {{\bf u}'} \right) {\gamma  \left( u' \right)} {\Phi \left( {\bf r} \right)}.
 \label{energysactoso}
\end{eqnarray}
  Similarly as in the space-time transformations we do not recover Lorentz symmetry between affected and unaffected observers. We only recover the SRT Lorentz transformations for energy-momentum for locally coincident   or similarly affected observers. In the $({{\bf p}},E)$-GMLT  the departure  from the Lorentzian symmetry in the noncoincident configuration of affected observers will be effective in causing the required gravitational dynamics. 

\subsection{GMLT metric and invariants}
The  $({{\bf p}},E)$-GMLT  and  $({{\bf x}}, t)$-GMLT differ only  in the  overall  $\Phi$ factor which is relatively counterposed.  The relatively inverse gravitational influence in space-time GMLT and momen\-tum-energy GMLT excludes a purely `metric' interpretation of this gravitation model.   While the space-time GMLT leads to the expression of invariant line element  $d {s'}^2$ and metric
\begin{eqnarray}
d {s'}^2 = {c '}^2 d {t'}^2 - d{{\bf x}'}^2  = {\Phi \left( r \right)}^{-2} 
\left( {c ({\bf r})}^2 dt^2-d{\bf x}^2  \right),    g_{\mu \nu} = \{ \Phi^{2}, - \Phi^{-2} {\bf 1} \},
\end{eqnarray} 
  the momentum-energy GMLT  Eq. (\ref{momentumsotosac}, \ref{energysotosac}) leads to a distinct  `momentum-energy metric' as well.
\begin{eqnarray}
{m_0'}^2 {c'}^4 = {E'}^2 -  {p'}^2 {c'}^2 = {\Phi \left( r \right)}^{-2} \left(E^2 - {c ({\bf r})}^2
p^2 \right),    g_{\mu \nu} = \{ \Phi^{-2}, - \Phi^{2} {\bf 1} \}.
\end{eqnarray} 
In this scalar model  physical quantities are related to adjusted GMLT's, depending on their dimensional
units. Consequently, all natural `constants' can not \emph{a priori} be considered unaffected by gravitation and could be covariant with a GMLT.  E.g., we observe this feature in the velocity of light $c$, but on the contrary not in Planck's constant $h$ or the electromagnetic fine structure constant $\alpha$.    The invariance of $\alpha$ is trivial due to its dimensionless nature (\footnote{The fine structure constant,  using SI units, is given by $\alpha = e^2/(4 \pi \epsilon_0 \hbar c)$. Since $c  = (\epsilon_0 \mu_0)^{-1/2}$,  the  invariance of $\alpha$  is due to permeability and permittivity transforming identically over GMLT. The respective units are   $[\mu] =  J s^2/(C^2 m)$ and $[\epsilon] = C^2/(J m)$. From the GMLT's we see  $J \to J' \Phi$,  $m \to m' \Phi$ and $s \to s' \Phi^{-1}$,  then $[\mu]/[\mu'] = [\epsilon]/[\epsilon']$. From the invariance of $\alpha$ no inference can be made about the  possible gravitational affecting of electric charge  $C \to C' \Phi^\eta$. On the other hand, when Gauss units are used, the fine structure constant is given by $\alpha = e^2/( \hbar c)$. Then  the invariance of $\alpha$ and $\hbar$ lead to  $e^2/c = {e'}^2/c'$, and from Eq. (\ref{VSL}) now follows $e' = e \Phi^{-1}$. \emph{I.e.} gravitational affecting of electric charge when using statcoulomb units.}), while  the invariance of Planck's constant follows from the combination of
$({{\bf p}},E)$-GMLT  and  $({\bf x},t)$-GMLT covarying quantities. The simplification is due to the contraposition of
the $\Phi$ factor in  $({\bf x},t)$-GMLT eqs. (\ref{spacesotosac}, \ref{timesotosac}) and  $({{\bf p}},E)$-GMLT
eqs. (\ref{momentumsotosac}, \ref{energysotosac}):
\begin{eqnarray}
{{\bf p}'}. d{\bf x}'  -  {E'}  d t'  &=& {{\bf p}} . d{\bf x}   -  E  dt. 
\end{eqnarray}
Taking into account the covariance of the Einstein-Compton relations for corpuscular light relative to the GMLT $S_0$ and $S'$:
\begin{eqnarray}
E = h \nu    \ , \  p = \frac{h}{\lambda}  & {\rm and}&   {E'} = h' \nu'    \ , \  p' = \frac{h'}{\lambda'},  \label{EinsteinCompton}
\end{eqnarray}
we have, with ${\bf k } ={\mbox{\boldmath{$\lambda$}}}/\lambda^2$, ${\bf k }' =
{\mbox{\boldmath{$\lambda$}}}'/{\lambda'}^2$
\begin{eqnarray}
 h' (  {\bf k }' . d{\bf x}'  -  \nu'  dt' )  &=& h (  {\bf k } . d{\bf x}  -  \nu  dt ). \label{phases}
\end{eqnarray}
Then, given the trivial invariance of a dimensionless phase, the GMLT-invariance of Planck's constant is obtained:
\begin{eqnarray}
h &=& h'.  \label{Planckinvariance}
\end{eqnarray}
The straightforward but  tacit conditions for the validity of this result are:  \emph{i)}   $\Delta \Phi \to 0$  when  $\Delta x \to \lambda$ for  Eq. (\ref{phases}), and foremostly \emph{ii)}  the adaptable parameter $\delta$ as required by  its fixing  Eq. (\ref{delta=0}).
In subsection Eq. (\ref{spectralshift}) we apply the invariance of Planck's constant Eq. (\ref{Planckinvariance}) in solving the problem of the spectral shift  of light in gravitational fields, requiring  and therefore endorsing the choice Eq. (\ref{delta=0}) for $\delta$. 

\subsection{Newtonian fit and static field equation \label{Newtonfit}}
The identification of $\Phi$ needs to be done in order to solve mechanical problems in practice. This will be done by comparing expressions of the energy change when a mass is lowered in a gravitational field according two affected observers: the first observer, {$S'_w$}, is at rest relative to the background field $\Phi$, and  the second observer, {$S'_e$}, is   the eigen observer of the lowered test mass.  In order not to have kinematical contributions we consider the test mass to be at at rest in the field at the start and end of the lowering. This will allow the static observer {$S'_w$} to identify the energy shift of the  mass as a pure Newtonian potential energy shift. The eigen observer $S'_e$  will invariably attribute ${{\bf p}'}_e=0$ and ${E'}_e = {m_0'}  {c '}^2$ both at the start and end of the lowering of the mass. 
In this static configuration the energy GMLT between {$S'_e$} and {$S'_w$} is simply: 
\begin{eqnarray}
   {E'}_w  \Phi_w  &=&  {E'}_e     \Phi_e. 
\end{eqnarray}
The lowering of the test body at {$S'_e$} leads to a change in static Newtonian potential energy $E'_w = U' (r') $ relative to {$S'_w$}, while  for {$S'_e$}   the energy remains ${E'}_e = {m_0'}{c '}^2$, thus:
\begin{eqnarray}
\Phi_w  d {E'}_w &=&  {E'}_e    d\Phi_e.  
\end{eqnarray} 
The  lowering of the mass in the end makes  {$S'_e$} and  {$S'_w$}  locally coincident, \emph{i.e.} $\Phi_w = \Phi_e$, then:
\begin{eqnarray}
\frac{d  \Phi_w}{\Phi_w}  &=&      \frac{1  }{{m_0'} {c '}^2} d \left(  - G' {m_0'} \int_{\rm S}   \frac{
\rho({\bf r}")  }{\vert {{\bf r}'} - {{\bf r}"} \vert} d^3 {r"} \right). 
\end{eqnarray} 
This directly solves to the expression: 
\begin{eqnarray}
  \Phi_w  ({\bf r}') &=& \exp \left[{ - \frac{ G'   }{ {c '}^2}  \int_S \frac{\rho({{\bf r}"})  }{\vert {{\bf r}'} - {{\bf r}"} \vert} d^3 {r}"}\right].  
 \label{generalPhiac} 
\end{eqnarray}
Then $\Phi_w$, specifically  outside a spherically  symmetric  source of mass $M'$, is given by: 
\begin{eqnarray}
\Phi_w ({r}')  \ = \ \exp \left[- \frac{ \kappa'   }{r'} \right]  &,&  \kappa' \ \equiv \ - \frac{G' M'}{{c '}^2},   \label{SSSPhiac}
\end{eqnarray}
where $\kappa'$  is  formally equal to half the (critical)  Schwarzschild radius.  {\emph A priori}, no singular features are expected in the affecting function $\Phi$, conform the gravitational scaling   of space and time intervals by $\Phi$, Eq. (\ref{staticrelation}). Evidently the accuracy of the Newtonian fit does not allow a physical interpretation of this closed exponential form, of which the valid expansion order in $\kappa$ should be verified in comparison with  GRT predictions.  When treating the GRT problems in the next Section we will require the expression $\Phi ({\bf r})$  from the perspective of $S_0$ instead of $S'_w$. The exact transcription of $\Phi_w ({\bf r}')$, Eq. (\ref{generalPhiac}), into the $S_0$ perspective requires a conventionally defined measurement protocol for extensive length. \\
If on the other hand  the unaffected observer $S_0$ attributes due the lowering of the test mass a \emph{scaled} Newtonian energy difference  equal to $d E = \Phi d U_N (r)$, then a field equation in $S_0$ perspective can be obtained. This energy shift attribution merely implies that for $S_0$ the Newtonian energy difference is affected identically as rest mass is by the gravitational field. In that case the Newtonian fitting procedure, with $d E = {m_0'} {c '}^2 d \Phi$,  from the perspective of  $S_0$ leads to:
\begin{eqnarray}
{m_0'} {c '}^2 d \Phi = \Phi d U_N (r),
\end{eqnarray}
and the field equation in $S_0$ perspective follows:
\begin{eqnarray} 
  \Delta  \Phi  \ = \ \frac{4 \pi G' }{{c '}^2} \rho({\bf r} ) \Phi  +  \frac{\left(\nabla \Phi \right)^2}{\Phi} , &&    \Phi  \ = \ \exp \left[ - \frac{ G'   }{ {c '}^2}  \int_S \frac{\rho({\bf r}^*)  }{\vert {\bf r} - {\bf r}^* \vert} d^3 r^*\right].  
 \label{generalPhi} 
\end{eqnarray}
For a spherically symmetric source of radius $R$ and mass $M$ this leads to:
\begin{eqnarray}
 \Phi ({\bf r})  \ = \ \exp \left[ - \frac{\kappa}{r} \right], &&  r > R , \ \ \   \kappa \equiv - \frac{G' M}{{c '}^2}.  \label{SSSPhi}
\end{eqnarray}
The affecting function $ \Phi ({\bf r})$ thus consistently satisfies a static gravitational field equation (\footnote{Based on invariance requirements, a similar equation for a point source was derived by Torgny Sj\"odin (private communication).  Equation (\ref{generalPhi}) resembles the static field equation proposed by Einstein in his precursor VSL-gravitation theory (\cite{Einstein19071912grouped}, {\bf 38}, 456,  Eq. (3b)): $ c \Delta c - \frac{1}{2} \left( \nabla c\right)^2 = k c^2 \rho $, with $c$ the velocity of light, $k$ the universal gravitational constant, and $\rho$ mass density (see also \cite{Norton1992FoPh}).}) with  source terms due material density $\rho({\bf r} )$ and the gravitational field energy density $\left(\nabla \Phi \right)^2/\Phi $.\\
Expression  (\ref{SSSPhi}) of $\Phi ({\bf r})$, as a scaling function, satisfies the conditions of  monotonous increase $\partial_r \Phi > 0 \ (r \neq 0)$, boundedness  $0\leq \Phi\leq 1$, and unaffected limit  $\Phi_\infty = 1$. While the closed exponential form $\Phi ({\bf r})$ is now sustained by the  field equation, its validity is still subject to  comparison with endorsed GRT predictions. In view of the closed form Eq. (\ref{SSSPhi}) ---appropriate for  gravitational scaling--- any critical phenomena  at $r = \kappa$ appears only due to truncation of order  in the coupling parameter G.  \\
In the preceding development we  have conservatively taken the mass  density $\rho$ as the primary source of the gravitational field in the Laplace equation.  In further development of the model, and if required by theoretical extension, supplementary source terms  like stress tensor $T_{\mu\mu}$ or electromagnetic energy density should be considered. Finally we mention once more the effect of the adaptable parameter $\delta$ when different from its fixed value Eq. (\ref{delta=0}): the procedure of the Newtonian fit with  parameter $\delta$ leads to  $ \Phi ({\bf r})  = \exp \left[ -  \kappa/(1-\delta) 1/r \right]$. Then the scaling boundary  conditions require $\delta <1$. We notice the  adaptable parameter changes the effective Schwarzschild radius, this feature is not  required in obtaining correct  predictions  of gravitational mechanics.
\subsection{Effective tensorial rank of interaction}
With the premisses for the scalar gravitational model established, we can now discuss  the  generally accepted `no-go' assessment of such models as mentioned in the introductory section, e.g. by  Cavalleri and Spinelli \cite{CavalleriSpinelli1983aFoPh}.
In order to compare the modified scalar model to classical field models we compare the Lagrangians in  the `minimal coupling' description (\cite{CavalleriSpinelli1983aFoPh},  Eq.(2)). Applying the  Legendre transformation on Eqs. (\ref{massHamiltonian}, \ref{lightHamiltonian}), gives the Lagrangian expressions:
 \begin{eqnarray}
L_{part} \equiv \mathbf{p}.\mathbf{v} - H_{part.} = - m_0'  \frac{\Phi}{\gamma(v)} &,& L_{phot.} \equiv \mathbf{p}.\mathbf{v}_c - H_{phot.} = 0. \label{lagrangians}
\end{eqnarray}
We do obtain the correct limit relation $v \to c$ between particles and  photons. But a perturbation development of the photon Lagrangian on $v_c$ is not possible $(v_c/c =1)$. For particles we obtain, using $\Phi=\exp\left[\varphi\right]$,   Eq. (\ref{SSSPhi}), the approximation:
\begin{eqnarray}
L_{part} &=& - m_0' {c'}^2 \left( 1+ \varphi+ \frac{1}{2}\varphi^2 \right)  \left( 1 -\frac{1}{2} \frac{v^2}{{c'}^2} \left( 1- 4 \varphi + 8 \varphi^2 \right) \right)  + O (\kappa^2, v^4/{c'}^4).
\end{eqnarray}
The components of  the interaction  and free  Lagrangian are easily identified.  We obtain, with $v^{0} \equiv 1$,  $L_{part} = L_{0} +L_{int.}$ and minding the hybrid form $\gamma (v/c')$, to first order in the coupling parameter (which is sufficient for the present purpose);
\begin{eqnarray}
L_0 &=& - \frac{m_0'   {c'}^2}{\gamma (v/c')}, \\
L_{int.} &=& - m_0'   {c'}^2  \psi_{\alpha \beta} \frac{v^\alpha v^\beta}{ {c'}^2}  + O (\kappa^2, \kappa v^4/{c'}^4) ,\ \ \ \ \ \ \psi_{\alpha \beta} =  \left\{ 1 , \frac{3}{2} \delta_{ij}\right\}\varphi, 
\end{eqnarray}
which is  the first-order part of the particle Lagrangian in GRT (\cite{Weinberg1972}, Eq. 9.2.3). In effect the minimal coupling description of the particle Lagrangian identifies the scalar model as rank-2 tensor model, with static source potential $\psi_{\alpha \beta} $. This behaviour comes about through $c(r)$ in the kinematical  $\gamma$-factor. The ratio $v^2/c^2$ systematically gives the appearance of  higher rank coupling; $v^2/c^2 =\Phi^{4}\delta_{ij}v_i v_j/{c'}^2$.  While the premisses for the present model clearly indicate we are considering a scalar gravitation field, the principle of gravitational affecting of quantities as mass and speed of light give the model a rank-2  tensor-like property. \\ 
For the description of  photons we  rely on the Hamiltonian scheme (or $\lim_{v \to c}$ of  Eq. (\ref{gvec})). In the next Section its  accordance with GRT and experiments is shown. The present  scalar model thus by--passes the limitation of the `classical' scalar field description in flat space time because essentially  it is of the  spatially-VSL type.

\section{Celestial mechanical GRT-experiments\label{exps}}
The  `basic' GRT experiments will be described in  {$S_0$} perspective;  {\it i)} the deflection of light by the sun, {\it ii)} the precession of orbital perihelia, {\it iii)} the gravitational delay of radar echo, and $S'$ perspective; {\it iv)} the gravitational redshift of spectral  lines.
The results can be transformed by $({\bf x}, t)$-GMLT Eqs. (\ref{spacesotosac}, \ref{timesotosac}) and  $({\bf p}, E)$-GMLT Eqs. (\ref{momentumsotosac}, \ref{energysotosac}) into the affected {$S'$} perspective when necessary.\\
Hamilton's principle will be applied from the perspective of the unaffected observer $S_0$. Evidently  the $E$-GMLT Eq. (\ref{energysotosac}) between $E$ and $E'$ impedes the simultaneous validity of the principle of energy conservation in both affected and unaffected perspectives in a straightforward manner. For energy conservation relative to $S_0$, a static affected observer $S'$ will attribute a change of energy $dE' /E' = - d \Phi /\Phi$. Note however that this asymmetrical feature will  resolve the problem of gravitational  redshift (see Section \ref{spectralshift}).\\
The Hamiltonian expression (\footnote{The corresponding  Lagrangian for particles (\ref{lagrangians}, a) is precisely the one  used by
Sj\"odin \cite{Sjodin1990FoPh}, and  the one proposed  by Einstein, $ - m \sqrt{c^2-q^2} $, in his early Lorentz-covariant gravitation model (\cite{Einstein19071912grouped}, {\bf 38},  {\em Nachtrag zur Korrektur}) })   Eq. (\ref{massHamiltonian}) for mass  and expression Eq. (\ref{lightHamiltonian}) for light (${m_0'} = 0$) are used:  
\begin{eqnarray} 
H    &=&   \sqrt{ {m_0 }^2  {c ({\bf r})}^4 + {c ({\bf r})}^2 p^2 }   \ = \  {\Phi \left( r \right)} \sqrt{  {m_0'}^2 {c '}^4       + p^2  {c '}^2{\Phi \left( r \right)}^2}. \label{Hamiltonians}
\end{eqnarray}
For a source with spherical symmetry, Eq. (\ref{SSSPhi}), the central force leads to
conservation of the angular momentum. The motion is constrained to the $(r, \varphi)$  plane, with $\theta= \pi/2$,  by the
Hamilton equations:
\begin{eqnarray}
& H    =  E_0, & \label{Hamiltonenergyconservation}\\
& p_\theta  =  p_{\theta_0}  = 0 \ \ , \ \  \dot{ \theta } \ = \   0,   & \\
& p_\varphi  =   p_{\varphi_0}    \ \ , \ \    \dot{ \varphi }\ = \  \frac{{c (r)}^2}{E_0 r^2}  {p_\varphi}_0, &
\label{Hamiltonangmomconservation}\\
&\dot{ p_r }  =  -  \left( E_0 + (p_r^2 +  \frac{p_{\varphi_0}^2}{r^2})  \frac{  {c (r)}^2}{ E_0 }    \right)
\frac{\partial_r {\Phi \left( r \right)}}{{\Phi \left( r \right)}}     + \frac{  {c (r)}^2}{E_0} \frac{p_{\varphi_0}^2}{r^3}   
\ \     , \ \    \dot{ r }  \ = \     \frac{{c (r)}^2}{  E_0}       p_r.   & \label{Hamiltonrad}
\end{eqnarray}
From which the  orbital equation of a test body in a spherical symmetrical field, with conserved quantities $E_0, p_{\theta_0},
p_{\varphi_0} $,  is obtained:
\begin{eqnarray}
\frac{d r }{d \varphi}   &=&   r 
\sqrt{ \frac{r^2}{{c (r)}^2   }\frac{\left(  E_0^2 -{m_0 }^2 {c (r)}^4
\right)}{p_{\varphi_0}^2} -   1}. \label{orbitalequation}
\end{eqnarray} 
The  expressions  for force {\bf f} on,  and  acceleration $ {\bf g}$ of, a test mass by a gravitational field are readily obtained as well:
\begin{eqnarray}
{\bf f}  &=&  - E_0 \left(2  - \frac{ 1}{ {\gamma  \left( v\right)}^2}    \right) \frac{\nabla {\Phi \left( r
\right)}}{{\Phi \left( r \right)}}  \label{fvec}\\
 {\bf g}  &=&   4  {\bf v}\frac{{\bf v}.\nabla {\Phi \left( r \right)}}{{\Phi \left( r \right)}}  - \left({c ({\bf
r})}^2+ v^2\right) \frac{\nabla {\Phi \left( r \right)}}{{\Phi \left( r \right)}}. \label{gvec}
\end{eqnarray}

\subsection{Deflection of light by a gravitational source} The unbound states of light are characterized by a single flexion
point ($\dot p_r = 0$) in the orbital at the point of minimal approach to the gravitational source. The flexion point and the impact parameter are used to characterize the light orbital. The angle $\alpha$  is defined between the radial vector ${\bf 1}_r$ and the tangent vector
${\bf 1}_t$ of the orbital, then: $ \tan \alpha = \frac{r d \varphi}{ d r} $. 
The impact parameter $b$ is obtained from  asymptotic orbital conditions: $ r_{\rm in} \sin \alpha_{\rm in} \vert_{r_{\rm
in} \to \infty} = b $. Setting the observation of deflection again asymptotically, $r_{\rm out} \to \infty$,  the
deflection angle $\alpha_{\rm D}$ is defined by:
\begin{eqnarray}
\alpha_{\rm D} &=& 2 \vert \varphi_\infty - \varphi_{{r_-}}\vert - \pi.
\end{eqnarray} 
The light orbital equation is obtained by putting ${m_0 }=0$ in  Eq. (\ref{orbitalequation}), and using the condition   $\dot p_{{r_-}} = 0$ at the point of shortest approach, then $ \frac{E_0}{p_{\varphi_0}} = \frac{{c '}}{b} =
\frac{{c_-}}{{r_-}}$:
\begin{eqnarray}
 \frac{d r }{ d \varphi} &=& 
 r  \sqrt{ \frac{{c_-}^2}{{c (r)}^2}\frac{r^2}{{r_-}^2} -   1} .
\label{lightorbitalequation}
\end{eqnarray}
The integration is evaluated in the approximation $O(\kappa^2/{r_-}^2)$
\begin{eqnarray}
  \varphi_\infty -\varphi_r  &\approx&
 \int_r^\infty \left(1 + 2 \kappa   \frac{r}{{r_-}}\frac{1}{r+{r_-}}\right)
\frac{d r}{r\sqrt{  r^2 {r_-}^{-2} -   1} }, \\
&\approx& \arcsin \frac{{r_-}}{r} + 2 \kappa \left(
\frac{1}{{r_-}} - \frac{\sqrt{r^2-{r_-}^2}}{{r_-}(r+{r_-})} \right).
\end{eqnarray} 
Then the deflection angle $\alpha_{\rm D}$ in  $O(\kappa^2/{r_-}^2)$ is given by:
\begin{eqnarray}
\alpha_{\rm D} &=& 2 \left\vert \frac{\pi}{2} + 2 \frac{\kappa}{{r_-}} \right\vert - \pi.
\end{eqnarray} 
This is the  GRT result in coordinate space-time:
\begin{eqnarray}
\alpha_{\rm D} &\approx& 4 \frac{G M}{{c '}^2 {r_-}},
\end{eqnarray} 
which is the right, observed value.

\subsection{The precession of orbital perihelia} 
The bound state of a massive test body has two flexion points, the perihelion $\frac{d r }{d \varphi} \vert_{r = r_-} = 0 $ and aphelion $\frac{d r }{d \varphi} \vert_{r = r_+} = 0 $ of Eq. (\ref{orbitalequation}),  in its orbit around a gravitational source.  The initial values $E_0$ and $p_{\varphi_0}$ can be expressed using the values at the extrema of the
bounded orbits:
\begin{eqnarray}
 E_0    \ = \   {m_0'} {c '}^2 \sqrt{  \frac{  {r_+}^2 {\Phi_+}^{-2}  -
 {r_-}^2 {\Phi_-}^{-2}}{   {r_+}^2 {\Phi_+}^{-4}  -
{r_-}^2{\Phi_-}^{-4}}}   &,&  p_{\varphi_0}   \ = \ {m_0'} {c '} \sqrt{ -  \frac{{\Phi_+}^2 - {\Phi_-}^2
}{ {\Phi_+}^4 {r_+}^{-2} -
{\Phi_-}^4{r_-}^{-2} }},  \\
{r_+} > {r_-} ,\  {\Phi_+} > {\Phi_-} &:&    \frac{{r_+} }{{\Phi_+} }  >   \frac{{r_-} }{{\Phi_-}} 
, \  \frac{{r_+}^2}{{\Phi_+}^4}  >   \frac{{r_-}^2}{{\Phi_-}^4},
\end{eqnarray} 
where the inequalities are obtained directly from $\frac{d r }{d \varphi} \vert_{r = r_\pm} = 0 $.
The orbital Eq. (\ref{orbitalequation}) can now be expressed using the extremal values:
\begin{eqnarray}
\hspace{-20 pt}\frac{d r }{d \varphi}  &=&  r^2  \sqrt{ \frac{1}{\Phi^4} \left( \frac{{\Phi_-}^4}{{r_-}^2}
\left(\frac{{\Phi_+}^2-\Phi^2}{{\Phi_+}^2-{\Phi_-}^2}\right) +
\frac{{\Phi_+}^4}{ {r_+}^2} \left(\frac{{\Phi_-}^2-\Phi^2}{{\Phi_-}^2-{\Phi_+}^2}\right)\right) -  
\frac{1 }{r^2}}. \label{integrand}
\end{eqnarray} The angular perihelion shift
$\Delta \varphi$ is defined by:
\begin{eqnarray}
\Delta \varphi &=& 2 \vert \varphi({r_+}) - \varphi({r_-}) \vert - 2 \pi.  
\end{eqnarray}
The integration of the equation requires some standard approximations and substitutions. 
We approximate  till $O(\kappa^3/r^3)$ the integrand according a quadratic form as integrand has zero points at $r =
r_-$ and $r = r_+$:
\begin{eqnarray}
\frac{1}{{\Phi \left( r \right)}^4} \left(\frac{f_-}{r_-^2} + \frac{f_+}{r_+^2} \right) - \frac{1}{r^2} &\approx& - \alpha 
\left(\frac{1}{r_- } - \frac{1}{r} \right) \left(\frac{1}{r_+ } - \frac{1}{r} \right), 
\end{eqnarray} 
with $\alpha$  a constant that can be evaluated by considering the value of the
integrand for $r \to \infty$:
\begin{eqnarray}
  \alpha  &=& - \frac{r_- r_+}{\Phi_\infty^4} \left(\frac{{f_-}_\infty}{r_-^2} + \frac{{f_+}_\infty}{r_+^2} \right). 
\end{eqnarray} 
With;
\begin{eqnarray}
  \Phi_\infty = 1  \ , \ {f_-}_\infty =  {\Phi_-}^4
\left(\frac{{\Phi_+}^2-1}{{\Phi_+}^2-{\Phi_-}^2}\right) \ , \ {f_+}_\infty =  {\Phi_+}^4  
\left(\frac{{\Phi_-}^2-1}{{\Phi_-}^2-{\Phi_+}^2}\right).
\end{eqnarray} 
Then; 
\begin{eqnarray}
  \alpha  &=& \frac{ r_+ {\Phi_-}^4 }{r_-}
\left(\frac{1-{\Phi_+}^2}{{\Phi_+}^2-{\Phi_-}^2}\right) - \frac{r_-  {\Phi_+}^4  }{r_+} 
\left(\frac{1- {\Phi_-}^2}{{\Phi_+}^2-{\Phi_-}^2}\right). \label{alpha}
\end{eqnarray} 
We now consider the integral of the orbital angle:
\begin{eqnarray}
\varphi (r) -\varphi(r_-) &\approx&  \frac{1}{\sqrt{\alpha}} \int_{r_-}^r \frac{d r}{ r^2
\sqrt{\left({r_-}^{-1}- {r}^{-1}\right)\left({r}^{-1}- {r_+}^{-1}\right)}}.
\end{eqnarray} 
This part of the integration is a standard (\cite{Weinberg1972}, sec. 8.6):
\begin{eqnarray}
\varphi (r_+) -\varphi(r_-) &\approx&  - \frac{\pi}{\sqrt{\alpha}}.
\end{eqnarray} 
The perihelion shift angle is then given by:
\begin{eqnarray}
\Delta \varphi   &\approx&  2 \pi \left(   \frac{1}{\sqrt{\alpha}} - 1\right). 
\end{eqnarray} 
Approximation of $\alpha^{- 1/2 }$ to order $O(\kappa^2/r^2)$, and $\Phi =
\exp \left[{- \kappa/r}\right]$ till $O(\kappa^3/r^3)$ gives:
\begin{eqnarray}
  \alpha^{-\frac{1}{2}}  &=&        1 +   \frac{3}{2}\kappa \left(\frac{1}{r_+}  +\frac{1}{r_-}\right) + O
\left(\frac{\kappa^2}{r^2} \right).
\end{eqnarray} 
Then the perihelion shift angle, up to $O \left( \kappa^2/r^2  \right) $,  is given by, 
\begin{eqnarray}
\Delta \varphi  &\approx &  \frac{6 \pi \kappa}{L},
\end{eqnarray} 
with $L \equiv1/2 \left( 1/r_+   + 1/r_- \right) $, i.e. the GRT result  in
coordinate space-time.\\
It is well known that in reality the found relativistic precession of Mercury is accompanied by a much larger precession of 532" per century. It is accounted for by the Newtonian interaction of remaining planetary sources (e.g. \cite{Weinberg1972}, Section III.9.5).
In order to reproduce this effect we should distinguish in the field Eq. (\ref{generalPhi}) the source density $\rho (r^*) =\rho_{Sol.} (r^*)+\rho_{rem.} (r^*)$ with components of the main, Solar, source and the remainder of planetary sources. This leads ---in time independent evolution approximation of the remainder planetary sources--- to $\Phi (\mathbf{r}) = \exp \left[{\phi(r)+\psi(\mathbf{r})}\right]$, with $\phi (r) = G' M_{Sol.}/({c'}^2  r)$ and  $\psi(\mathbf{r}) = \sum_i  G' m_i/ ({c'}^2   \vert \mathbf{r} -\mathbf{r}_i\vert)$. This addition of effects results, as in GRT (\cite{Weinberg1972},  9.4.11, 9.1.64),  to an additional potential term in the gradient term of the acceleration to first approximation:
\begin{eqnarray}
{\bf g}  &=&   - {c'}^2 \nabla \left({\phi ( r ) + 2  \phi ( r ) ^2+ \psi(\mathbf{r})} \right)  +  4  {\bf v}{\bf v}.\nabla {\phi ( r )}   -   v^2  \nabla {\phi ( r )}. 
\end{eqnarray}
This acceleration expression corresponds only to the fixed field approximation of GRT (\cite{Weinberg1972}, 9.2.1), but does account correctly for the precession effect due the remainder Newtonian planetary sources. We refer to \cite{Broekaert2004a} for a more realistic ---in terms of the true planetary system--- time-dependent formulation of source movement in the framework of the present model.

\subsection{ The gravitational delay of radar echo} 
The travel time of light is obtained from the variable velocity of light at every point of its orbit between source and
observer.  From the Hamiltonian equation (\ref{Hamiltonrad}, b), with conserved energy $E_0= p c $
Eq. (\ref{Hamiltonenergyconservation}) and angular momentum  $p_{\varphi_0}$ Eq. (\ref{Hamiltonangmomconservation})  we
obtain:
\begin{eqnarray}
\dot{ r } &=&    {c (r)}  \sqrt{1- \frac{p_{\varphi_0}^2}{E_0^2 }\frac{ {c (r)}^2}{ r^2}}.     
\end{eqnarray} The radial velocity vanishes at the point of closest approach near the Sun, $r={r_-}$, i.e. $\dot r
\vert_{r={r_-}} = 0$ giving $  p_{\varphi_0}/E_0 =      {r_-}/ {c_-}  $.
Then, light traveling over its orbit from $r_i$ to $r_f$ takes the time $\Delta t$:
\begin{eqnarray}
\Delta t  &=&  \int_{r_i}^{r_f} \frac{dr}{c (r)  \sqrt{1-    {r_-}^2 r^{-2} {c_-}^{-2} 
{c (r)}^2 }}.
\end{eqnarray}  
The travel time $\Delta t (r, {r_-})$, approximated till $ O \left(\frac{\kappa^2}{r^2}
\right)$, is found to be:
\begin{eqnarray}
\Delta t (r, {r_-}) &\approx& \frac{1}{{c '}} \int_{r_-}^r  \left(1 + 2 \frac{\kappa}{r}
 +  \frac{\kappa {r_-}}{r (r+{r_-})}\right)\frac{1}{\sqrt{1-  {r_-}^2  r^{-2}}} dr.
\end{eqnarray} 
The three components of the integrand  lead respectively to:
\begin{eqnarray}
\Delta t (r, {r_-})  \hspace{-7 pt}&\approx&\hspace{-7 pt}   \frac{\sqrt{r^2 -  {r_-}^2 } }{{c '}} + \frac{\kappa }{{c '}} \left( 2 \ln  \frac{r+\sqrt{r^2
-{r_-}^2}}{{r_-}}
 +  \left(\frac{r- {r_-}}{r+{r_-}} \right)^{\frac{1}{2}}\right).  
\end{eqnarray}
i.e. the delay term of GRT in coordinate time (\cite{Weinberg1972}, equation 8.7.4).

\subsection{Gravitational shift of spectral lines \label{spectralshift}} 
The effect of gravitation on the frequency of light over its orbit is calculated using
 the Hamiltonian description Eq. (\ref{Hamiltonians}) and the Einstein-Compton relations  for corpuscular light Eq. (\ref{EinsteinCompton}). 
In the unaffected perspective of $S_0$ the conservation of energy $E=E_0$, 
Eq. (\ref{Hamiltonenergyconservation}), is satisfied on the light orbital. 
Accordingly the unaffected observers $S_0$ attributes to light the frequency and wavelength:
\begin{eqnarray}
 \dot E = 0  &\to&     h \dot  \nu = 0  \ , \  \nu = \nu_0, \\
& &     \frac{ \dot \lambda}{\lambda} =      \frac{ \dot c}{c} \ , \  \lambda  =  \lambda_0 \frac{ \Phi^2}{\Phi_0^2}.  
\end{eqnarray}
The constancy of the light frequency relative to {$S_0$}  is the trivial result of energy conservation.  The  variation of wave length of light relative to $S_0$ is proportional to $\Phi^2$. Therefore {$S_0$} will attribute a `wave-length spectral shift':
 e.g. for  light emitted by a source at $r_s$ and observed at $r_o$, $r_o >> r_s $: 
\begin{eqnarray}
 \frac{\lambda_s - \lambda_o }{ \lambda_o}  &=&  -  2    \frac{\kappa}{r_s} + O \left(\frac{\kappa}{r_o}, \frac{\kappa^2}{r_s^2}.
 \right)
\end{eqnarray} 
Therefore  {$S_0$}   observes not a shift of frequency,  but a `red shift of the wavelength'  twice the magnitude an affected observer is expected to measure.\\
In the present scalar gravitation model the correct gravitational spectral shift of light is restored in the perspective of  affected
observers $S'$ only.     
With the GMLT-invariance of Planck's constant, Eq. (\ref{Planckinvariance}),  the observations of $S_0$ can be transformed into affected observations of $S'$ by either $({\bf x}, t)$-GMLT Eqs. (\ref{spacesotosac}, \ref{timesotosac}) or $({\bf p}, E)$-GMLT Eqs. (\ref{momentumsotosac}, \ref{energysotosac}). Both  GMLT's lead, for  ${\bf u}' = 0$, to the  $S_0$ to $S'$ relation for frequency and wave length: 
\begin{eqnarray}
\nu'  &=& \frac{ \nu}{ {\Phi \left( r \right)}}, \\
\lambda'   &=& \frac{ \lambda}{ {\Phi \left( r \right)}}. 
\end{eqnarray}
The affected observer $S'$ thus attributes to light, traveling from a source at $r_s$ to position  $r_o$ with $r_o >> r_s$, a frequency and wave length shift  equal to:
\begin{eqnarray}
  \frac{\nu'_{s}- \nu'_o}{\nu'_o }   &=&   \frac{\Phi (r_o)}{\Phi (r_s)} -1  \ =  \   \frac{\kappa}{r_s}  + O \left(\frac{\kappa}{r_o}, \frac{\kappa^2}{r_s^2} \right), \\
\frac{\lambda'_s-  \lambda'_o}{ \lambda'_o} &=&  \frac{ \Phi (r_s)}{  \Phi (r_o)} - 1 \ = \   -    \frac{\kappa}{r_s} + O \left(\frac{\kappa}{r_o}, \frac{\kappa^2}{r_s^2} \right).
\end{eqnarray} 
The gravitational spectral red shift as predicted by GRT is  recovered by the affected observer {$S'$}.  The   correct  frequency  and wave length spectral shift  relative to a static $S'$, as compared to $S_0$, is of course  due to the nonconservation of energy relative to {$S'$}.

\section{Conclusions} 
We presented the foundations of a scalar Lorentz-covariant gravity model based on alternative metrization by  consistent scaling of physical quantities in line with  Poincar\'e's geometric conventionalism and Lorentz-Poincar\'e type interpretation.  Two levels of description are related, the gravitationally affected observer (curved metric) and gravitationally unaffected observer (flat metric), by appropriate gravitationally modified Lorentz transformations or GMLT's. The inherent gravitational scaling function is nonsingular at any finite distance to a point source, while it is mathematically characterized by the Schwarzschild radius.\\
The model assumes an intrinsic spatially variable speed of light, depending on the gravitational field, but the GMLT's assure its observed local invariance. The presence of the variable speed of light, in the term $v^2/c^2$ in the Lagrangian, lends  the model rank-2 tensorial  properties. In this sense our `modified' scalar model by-passes the critique on `standard' scalar models in flat space \cite{CavalleriSpinelli1983aFoPh}. The Hamiltonian  for a test particle  in a gravitational field obtained by energy-momentum GMLT expresses an essential effect according Mach's mass induction principle. Developed for particles and photons in a static spherically symmetric gravitation field, it reproduces till required Post-Newtonian order the results for the four `basic' experiments of General Relativity Theory.\\ The  here presented work focused the simpler  static spherically symmetric configuration with test masses only and thus  ignored effects, e.g. due to:  size and proper kinematics of sources and gravitating objects,  or gravitational waves.  Some of the latter extensions of configuration will be covered in upcoming work concerning this scalar model \cite{Broekaert2004a, Broekaert2004b}. 

\section*{Acknowledgments}
The author wishes to thank,  Dr. T. Sj\"odin for the many enlightening discussions on his  Lorentz-Poincar\'e approach to gravitation, Prof. M. Arminjon for  rigorous reviewing and advise on interpretations, and Prof. G. Cavalleri for the valuable suggestion of relating the model to    gravitational field theory and supplying pertinent references.

\appendix

\section{Appendix}
\subsection*{Space and Time GMLT between affected observers}
The space time GMLT of the second type, \emph{i.e.} between  two distinctly, gravitationally and kinematically, affected observers $S'_1$ and $S'_2$, $\Phi_1 \neq \Phi_2$,  ${\mathbf u_1} \not  \parallel {\mathbf u_2}$  is obtained by elimination of the {$S_0$} perspective from a composition of two GMLT's (\ref{spacesactoso}, \ref{timesactoso})  and   (\ref{spacesotosac}, \ref{timesotosac}).  This relation should exclude specific kinematic reference to the perspective of $S_0$ in order not to obtain referenceless kinematical expressions in the homogenous case  $\Phi_1 = \Phi_2$ . The second type GMLT, exclusively in terms of kinematical quantities relative to  {$S'_1$} and {$S'_2$}, are given by :
\begin{eqnarray}
d{\bf x}' _1 &=& {d{\bf x}'_2}_{\perp  \{12, 21\}}  \frac{\Phi_2}{\Phi_1} + \Gamma_{21}  \frac{\Phi_1}{\Phi_2}  {\bf u}'_{21}  d t '_2   -  \Gamma_{21}  \frac{\Phi_1}{\Phi_2}  \frac{d{\bf x}'_2.  {\bf u}'_{12}}{{u'_{12}}^2}  {\bf u}'_{21}  + \nonumber\\
 &  &    \hspace{-13 pt}\frac{d{\bf x}'_2.  {\bf u}'_{21_{\perp  12}}}{{u'_{21_{\perp 12}}}^2}  \frac{\Phi_2}{\Phi_1} \left[{\bf u}'_{21}\left(1 + A \frac{\Phi_1^4}{\Phi_2^4}  \left(\frac{1}{\Gamma_{12}}-1\right)- 
B  \frac{\Phi_1^2}{\Phi_2^2} \Gamma_{21} \left( 1 +\frac{\gamma_1}{ \Gamma_{12}}\right) 
 \right) \right.    \nonumber \\
& &\hspace{35 pt}\left.   + {\bf u}'_{12}\left( \cos T \left\vert\frac{u'_{21}}{u'_{12}}  \right\vert + A  \frac{\Phi_1^2}{\Phi_2^2}\left(1-\frac{1}{\Gamma_{21}}\right) - B \left(1 + \gamma_2\right)\right)\right],  \label{spacesactosac} \\ 
d t' _1& =&  \left( d t'_2 -  \frac{d{\bf x}'_2 . {\bf w}'_{12}}{{c'}^2}  \right) \Gamma_{21}  \frac{\Phi_1}{\Phi_2},  \label{timesactosac}
\end{eqnarray}
with:
\begin{eqnarray}
A \ = \ \frac{{u'_{21_{\perp 12}}}^2}{{c'}^2} \frac{(1+\gamma_1)(1+\gamma_2)}{\Gamma_{12} D^2} , \ \ B\ =\ \frac{{\bf u}'_{21_{\perp 12}}.  {\bf w}'_{12}}{{c'}^2}  \frac{1}{D},  \\
D \ = \ - \frac{(1+\gamma_1)(1+\gamma_2)}{\Gamma_{12}\Gamma_{21}} + \left( 1+\frac{\gamma_1}{\Gamma_{12}}\right) \left( 1+\frac{\gamma_2}{\Gamma_{21}}\right),
\end{eqnarray}
where the gamma factors $\gamma_1$ and $\gamma_2$ are expressed in terms relative to  {$S'_1$} and  {$S'_2$} according: 
\begin{eqnarray}
\gamma_1 & = & \sqrt{\frac{\Gamma_{12}^2 \gamma_{12}^{-2} - \Phi_1^4\Phi_2^{-4}}{1 -  \Phi_1^4 \Phi_2^{-4}}} ,\ \ \ 
\gamma_2 \ = \ \sqrt{\frac{\Gamma_{21}^2 \gamma_{21}^{-2} - \Phi_2^4 \Phi_1^{-4}}{1 -  \Phi_2^4 \Phi_1^{-4}}}, \label{gammaeenengammatwee}
\end{eqnarray}
and in which the `gamma' factors for kinematical contraction and dilation are:
\begin{eqnarray}
\Gamma_{21} \equiv \frac{\Phi_2}{\Phi_1}\left. \frac{d t'_1}{d t '_2}\right\vert_{d{\bf x}'_{2}=0} &=& \gamma_1\gamma_2 \left(1 - \frac{{\bf u}'_{1}.{\bf u}'_{2}}{{c'}^2} \frac{\Phi_2^2}{\Phi_1^2}  \right), \\
\Gamma_{12} \equiv \frac{\Phi_1}{\Phi_2}\left. \frac{d t'_2}{d t '_1}\right\vert_{d{\bf x}'_{1}=0} &=&\gamma_1\gamma_2 \left(1 - \frac{{\bf u}'_{1}.{\bf u}'_{2}}{{c'}^2} \frac{\Phi_1^2}{\Phi_2^2}  \right),
\end{eqnarray}
and where the relative frame velocities are:
\begin{eqnarray}
{\bf u}'_{12} \equiv  \left. \frac{d {\bf x}'_2}{d t '_2}\right\vert_{d{\bf x}'_{1}=0}  &=& \frac{ {\bf u}'_{2} - \left( {\bf u}'_{1\parallel 2} +  {\bf u}'_{1\perp 2} {\gamma_2}^{-1}\right)  \Phi_1^2{\Phi_2}^{-2} }{\left(1 -  {\bf u}'_{1}.{\bf u}'_{2} {c'}^{-2}  \Phi_1^2 {\Phi_2}^{-2}  \right)}, \label{uaceentwee} \\
{\bf u}'_{21} \equiv  \left. \frac{d {\bf x}'_1}{d t '_1}\right\vert_{d{\bf x}'_{2}=0}  &=& \frac{ {\bf u}'_{1} - \left( {\bf u}'_{2\parallel 1} + {\bf u}'_{2\perp 1}{\gamma_1}^{-1}\right)  \Phi_2^2{\Phi_1}^{-2} }{\left(1 - {\bf u}'_{1}.{\bf u}'_{2} {c'}^{-2} \Phi_2^2 {\Phi_1}^{-2}  \right)}, \label{uactweeeen}
\end{eqnarray}
and the `counter-scaled' relative frame velocities are:
\begin{eqnarray}
{\bf w}'_{12}    \equiv   \frac{ {\bf u}'_{2}  - \left( {\bf u}'_{1\parallel 2}  +   {\bf u}'_{1\perp 2} {\gamma_2}^{-1}
\right) \Phi_2^2{ \Phi_1}^{-2} }{1 -   {\bf u}'_{1} .  {\bf u}'_{ 2} {c'}^{-2}
 \Phi_2^2{\Phi_1}^{-2}}, \ \ 
{\bf w}'_{21}    \equiv  \frac{ {\bf u}'_{1} - \left( {\bf u}'_{2\parallel 1} + {\bf u}'_{2\perp 1} {\gamma_1}^{-1}\right)  \Phi_1^2 {\Phi_2}^{-2} }{\left(1 - {\bf u}'_{1}.{\bf u}'_{2} {c'}^{-2} \Phi_1^2 {\Phi_2}^{-2}  \right)},
\end{eqnarray}
and where  the `Thomas angle' is given by:
\begin{eqnarray}
\cos T &=& - \frac{{\bf u}'_{12}.{\bf u}'_{21}}{u'_{12} u'_{21}}.
\end{eqnarray}
The precise mathematical group symmetry of the modified Lorentz transformations (\ref{spacesactosac}, \ref{timesactosac}) is not apparent. Only  the space intervals orthogonal to  the kinematical  plane $\{{\bf u}'_{12}, {\bf u}'_{21}\}$ follow a simple scaling transformation. In the kinematical  plane  the analysis of the transformations into a  boost  and Thomas rotation is not recovered, which is  clear from the difference in  amplitude of $u'_{12}$ and $u'_{21}$, Eqs. (\ref{uaceentwee}, \ref{uactweeeen}). \\
Only in the homogeneous case,  when the Lorentz symmetry is recovered due $\Phi_1=\Phi_2$ and  $\Gamma_{21} = \Gamma_{12}  = \Gamma$, do the coefficients in the transformation correspond to trigonometric functions of the  Thomas angle  $T$:
\begin{eqnarray}
\left. \cos T \right\vert_{\Phi_1=\Phi_2} &=& \frac{\left(1 + \Gamma + \gamma_1 + \gamma_2\right)^2}{(1+\Gamma) (1+\gamma_1) (1+\gamma_2)} - 1.
\end{eqnarray}
In the general case the kinematical affecting factors do not correspond to the standard SRT form of $\gamma_{21} \equiv (1-{u'_{21}}^2/{c'}^2)^{-1/2}$ and $\gamma_{12} \equiv (1-{u'_{12}}^2/{c'}^2)^{-1/2}$. Their general relation is:
\begin{eqnarray}
\gamma_{21} &=& \Gamma_{21} \frac{\gamma_2^*}{\gamma_2} , \ \ \ \ \gamma_2^* =  (1-{u'_2}^2 {c'}^{-2}  \Phi_2^4\Phi_1^{-4} )^{-1/2}, \\
\gamma_{12} &=& \Gamma_{12} \frac{\gamma_1^*}{\gamma_1} , \ \ \ \ \gamma_1^* =  \left(1-{u'_1}^2 {c'}^{-2}  \Phi_1^4 \Phi_2^{-4} \right)^{-1/2}.
\end{eqnarray}
Some modified relations of SRT are valid in the general inhomogeneous  case:
\begin{eqnarray}
 {\bf w}'_{21}. {\bf u}'_{21} &=&  {\bf w}'_{12} .{\bf u}'_{12}, \\
\frac{1}{\Gamma_{12} \Gamma_{21}} &=& 1 - \frac{ {\bf w}'_{12} .{\bf u}'_{12}}{{c'}^2}.
\end{eqnarray}
The frame velocities of $S'_1$ and $S'_2$ relative to $S_0$, or the gravitational field $\Phi$, can be related to  quantities relative to $S'_1$ and $S'_2$ by inverting Eqs. (\ref{uaceentwee}, \ref{uactweeeen}):
\begin{eqnarray}
 {\bf u}'_{2} &=& \left({\bf u}'_{21} \frac{\gamma_1}{ \Gamma_{12}}\frac{\Phi_1^2}{\Phi_2^2}+  {\bf u}'_{12} \frac{1+\gamma_2/\Gamma_{21}}{1+1/\gamma_1} \right)\frac{ 1}{D} \left(1+\frac{1}{\gamma_1}\right)\left(1+\frac{1}{\gamma_2}\right),\\
 {\bf u}'_{1} &=& \left({\bf u}'_{12} \frac{\gamma_2}{ \Gamma_{21}}\frac{\Phi_2^2}{\Phi_1^2}+  {\bf u}'_{21} \frac{1+\gamma_1/\Gamma_{12}}{1+1/\gamma_2} \right)\frac{ 1}{D} \left(1+\frac{1}{\gamma_1}\right)\left(1+\frac{1}{\gamma_2}\right).
\end{eqnarray}
\subsection*{Energy and Momentum GMLT between affected observers}
Due the similar kinematic structure and the inverse overall affecting factor of $({{\bf p}},E)$-GMLT  (\ref{momentumsactoso}, \ref{energysactoso}) and $({\bf x},t)$-GMLT (\ref{spacesactoso}, \ref{timesactoso}),  the general energy-momentum transformation between affected observers can be deduced from their space-time relation (if $\Phi_i \to 1/\Phi_i$, then $\Gamma_{21} \to \Gamma_{12}$, ${\bf u}'_{12} \to {\bf w}'_{12} $, ...  while $\gamma_i$ remain invariant). 
The most general energy-momentum GMLT between two distinctly  affected observers $S'_1$ and $S'_2$, $\Phi_1 \neq \Phi_2$,  ${\mathbf u_1} \not  \parallel {\mathbf u_2}$ is then given by:
\begin{eqnarray}
{\bf p}' _1&=&  {{\bf p}'_2}_{\perp  \{12, 21\}}  \frac{\Phi_1}{\Phi_2} + \Gamma_{12}  \frac{\Phi_2}{\Phi_1}  {\bf w}'_{21}   \frac{E '_2}{{c'}^2}   -  \Gamma_{12}  \frac{\Phi_2}{\Phi_1}  \frac{{\bf p}'_2.  {\bf w}'_{12}}{{w'_{12}}^2}  {\bf w}'_{21}  + \nonumber\\
& & \hspace{-12 pt}\frac{{\bf p}'_2.  {\bf w}'_{21_{\perp  12}}}{{w'_{21_{\perp 12}}}^2} \frac{\Phi_1}{\Phi_2}  \left[{\bf w}'_{21} \left(1 +\tilde{A}\frac{\Phi_2^4}{\Phi_1^4}  \left(\frac{1}{\Gamma_{21}}-1\right)- \tilde{B} \frac{\Phi_2^2}{\Phi_1^2}  \Gamma_{12} \left(1 +\frac{\gamma_1}{\Gamma_{21}}\right)
 \right) \right.  \nonumber \\
& &\hspace{25 pt}  \left.  +{\bf w}'_{12}\left( \cos \tilde{T} \left\vert\frac{w'_{21}}{w'_{12}}  \right\vert +\tilde{A} \frac{\Phi_2^2}{\Phi_1^2}\left(1-\frac{1}{\Gamma_{12}}\right) - \tilde{B} \left(1 + \gamma_2\right)\right)\right], \label{momentumsactosac} \\ 
 E' _1&=&   \left( E'_2 -  {\bf p}'_2 . {\bf u}'_{12}  \right)  \Gamma_{12}   \frac{\Phi_2}{\Phi_1},  \label{energysactosac}
\end{eqnarray}
where ${\bf w}'_{21_{\perp 12}}$ now indicates the part of ${\bf w}'_{21}$ orthogonal and coplanar with ${\bf w}'_{12}$, and:
\begin{eqnarray}
&\tilde{A} = \frac{{w'_{21_{\perp 12}}}^2}{{c'}^2} \frac{(1+\gamma_1)(1+\gamma_2)}{\Gamma_{21} \tilde{D}^2}, \ \ 
\tilde{B} =  \frac{{\bf w}'_{21_{\perp 12}}.  {\bf u}'_{12}}{{c'}^2}  \frac{1 }{\tilde{D}},  \\
&\tilde{D} = - \frac{(1+\gamma_1)(1+\gamma_2)}{\Gamma_{12}\Gamma_{21}} + \left( 1+\frac{\gamma_1}{\Gamma_{21}}\right) \left( 1+\frac{\gamma_2}{\Gamma_{12}}\right), \\
&\cos \tilde{T} = - \frac{{\bf w}'_{12}.{\bf w}'_{21}}{w'_{12} w'_{21}}.  
\end{eqnarray}
Let us for example consider the configuration with $S_2'$ kinematically coincident with the test body, \emph{i.e.} ${\bf p}'_2 = 0$. Then energy and momentum attributed by $S'_1$ relate to the quantities in the self frame $S'_2$, with $E '_2 = m_0' {c'}^2$, according:
\begin{eqnarray}
{\bf p}' _1&=&   \Gamma_{12}    \frac{\Phi_2}{\Phi_1}   m_0' {\bf w}'_{21},  \\ 
 E' _1&=&  \Gamma_{12}      \frac{\Phi_2}{\Phi_1}  m_0' {c'}^2.  \label{selfenergysactosac}
\end{eqnarray}
While the velocity of $S_2'$ relative to $S_1'$ is $ {\bf u}'_{21}$, we obtain  an  expression   for  momentum  ${\bf p}' _1$ along the  counter-scaled velocity ${\bf w}'_{21}$. Thus dismissing, in this model, the conventional momentum definition in the  gravitationally affected perspective.  

\bibliography{ArxivGMLT4basicV2.bib}

\end{document}